\documentclass{aa}
\usepackage{epsf}
\newcommand{\etal}{{et al}\/.}
\newcommand{\new}[1]{{#1}}
\begin{document}
\headnote{Research Note}
\title{$B_{\rm gg}$ revisited: The environments of low-excitation
  radio galaxies and unified models}
\titlerunning{$B_{\rm gg}$ revisited}
\author{Martin J. Hardcastle}
\institute{Department of Physics, University of Bristol, Tyndall
Avenue, Bristol BS8 1TL, UK (m.hardcastle@bristol.ac.uk)}
\date{Version of \today}

\abstract{Recent measurements of the galaxy clustering environments
  around intermediate-redshift radio sources have suggested a
  systematic environmental difference between radio galaxies and
  radio-loud quasars, in contradiction to the predictions of simple
  unified models for the two classes of object. I show that the
  apparent difference arises mainly as a result of the properties of
  low-excitation radio galaxies included in the radio-galaxy sample,
  which tend to lie in significantly richer environments. The
  environmental properties of high-excitation radio galaxies and
  quasars are statistically consistent in the redshift range $0.15 < z
  <0.4$, as unified models would predict.  }

\maketitle

\section{Introduction}

In unified models for powerful radio galaxies and radio-loud quasars
(Scheuer 1987, Barthel 1987, 1989) isotropic properties of the two
classes of objects, such as their clustering environments, should be
statistically identical. However, the classical unified models work
best at high redshift (there are few low-redshift objects classified
as radio-loud quasars) and at these redshifts the clustering
environments of the sources are hard to measure. Some attempts have
been made to test the predictions of unified models using X-ray
observations of extended emission, assumed to be cluster-related
(Hardcastle \& Worrall 1999) but sample sizes are generally small and
{\it ROSAT}-based work is generally not reliable because of the
contamination of the X-ray emission by AGN- and radio-source-related
components. {\it Chandra} should provide significantly better
constraints from the X-ray once sufficient objects are in the archive,
though it will be important to take into account the significant
evolution undergone by the ICM of a radio-source host cluster between
$z \sim 1$ and 0. In the meantime, there is still an important r\^ole
for the traditional method of calculating the amplitude of the
galaxy-galaxy or quasar-galaxy spatial covariance function, hereafter
referred to as $B_{\rm gg}$ (e.g. Longair \& Seldner 1979; Yee \&
Green 1987; Prestage \& Peacock 1988; Yates, Miller \& Peacock 1989;
Ellingson, Yee \& Green 1991; Hill \& Lilly 1991; Yee \& Ellingson
1993). Although it is recognised that selection in more than one
optical filter is desirable to achive good statistics at high
redshifts (Barr \etal\ 2003), at lower redshifts simple $B_{\rm gg}$
determinations seem likely to be able to provide a statistical measure
of the environments of different classes of source.

Recently Harvanek \etal\ (2001: hereafter HESR) have carried out a
large systematic study of the clustering properties of 3CR (Spinrad
\etal\ 1985) radio sources. By making new observations, and collating
and cross-calibrating data from the literature, they were able to
obtain a nearly complete set of $B_{\rm gg}$ values in the redshift
range $0.15 < z < 0.65$. 3CR objects in this redshift range are almost
all luminous FRII-type radio galaxies or lobe- or core-dominated
quasars. Yee \& Green (1987) had shown that few low-redshift,
radio-loud quasars are located in rich environments. HESR aimed to
test the unified-model prediction that radio galaxies at these
redshifts would also tend to lie in poor environments. They obtained
the surprising result that the $B_{\rm gg}$ values of their 3CR radio
galaxies were not consistent with those of a sample of radio-loud
quasars compiled by Yee \& Ellingson (1993: hereafter YE93) in the
redshift range $0.15 < z < 0.4$, in the sense that a number of the
radio galaxies were found in significantly richer environments than
any quasar. An immediate objection to this result is that the YE93
quasars were not in general as radio-luminous in extended emission as
the 3CR comparison sample (otherwise they would all have been 3CR
objects themselves on the basis of their low-frequency extended
emission, whereas only a fraction of them are in fact in 3CR). The
effects of this bias are hard to quantify, but since (for equal jet
powers and source ages) simple radio-galaxy physics suggests that a
more luminous object will lie in a richer environment, it might tend
to produce a difference in the sense of the one observed by HESR. In
what follows I shall ignore this potential problem and concentrate
instead on the effects on HESR's conclusions of using a more
sophisticated variant of the unified model.

\section{Unification at low redshifts and the low-excitation objects}
\label{unif}

At low redshifts (or, more generally, low luminosities) the simple
unified model of Barthel (1989) breaks down, most obviously because of
the absence of low-redshift FRII quasars. I have argued elsewhere
(Hardcastle \etal\ 1998) that the broad-line radio galaxies (BLRG)
take the place of quasars at low redshifts/luminosities, although that
does not rule out the possibility that more luminous BLRG are objects
intermediate in viewing angle between quasars and narrow-line radio
galaxies (NLRG) (e.g. Dennett-Thorpe \etal\ 2000). However,
low-excitation radio galaxies (LERG) (Hine \& Longair 1979; Laing
\etal\ 1994), which are a significant population at low FRII
luminosities, cannot be unified with either BLRG or quasars if the
narrow-line emission is isotropic. Instead, it has been suggested
(Barthel 1994, Laing \etal\ 1994), that they form a separate
population, possibly unified at small angles to the line of sight with
FRII-like BL Lac objects. They have other anomalous properties:
Hardcastle \etal\ (1998) noted that the LERG in their $z<0.3$ 3CR/3CRR
sample were on average significantly smaller than the high-excitation
objects, and that they included a number of objects with prominent
jets and relaxed or absent hotspot structure. Hardcastle \& Worrall
(1999) found that the LERG in their {\it ROSAT}-observed subsample of
3CRR tended to have relatively luminous X-ray environments, whereas at
least some of the NLRG at similar redshifts were in poorer
environments. This motivates a re-analysis of the results of HESR
taking into account the emission-line type of the radio galaxies.

\begin{table}
\caption{Supplementary emission-line classifications for HESR sources}
\label{emlin}
\begin{center}
\begin{tabular}{llll}
\hline
Source&\multicolumn{2}{c}{Emission line classification}&Reference\\
&in HESR&adopted\\
\hline
3C\,28&--&E&1\\
3C\,48&--&B&2\\
3C\,49&--&N&3\\
3C\,67&--&B&4\\
3C\,93.1&--&N&3\\
3C\,99&--&N&3\\
\new{3C\,196.1}&--&N&3\\
3C\,234&N/B?&N&5\\
3C\,268.3&--&B?&9\\
3C\,277.1&--&B&6\\
3C\,288&--&E&7\\
3C\,295&--&N&8\\
3C\,299&--&N&3\\
3C\,303.1&--&N&3\\
3C\,346&--&E&3\\
3C\,455&--&B&6\\
\hline
\end{tabular}
\end{center}
\vskip 5pt
\begin{minipage}{\linewidth}
Emission-line types here are E (low-excitation), N (high-excitation,
narrow-line) and B (broad-line). References are as follows: (1) Schmidt
(1965) (2) Greenstein \& Schmidt (1964) (3) Spinrad \etal\ (1985) (4)
Laing \etal\ (1994) (5) See Hardcastle \etal\ (1997) for
the argument for assigning this to the N class (6) Quasar, assumed
broad-line (7) Smith \& Spinrad (1980) (8) Minkowski (1960) (9) Laing
\& Riley, in preparation.
\end{minipage}
\end{table}

\begin{figure}
\epsfxsize\linewidth
\epsfbox{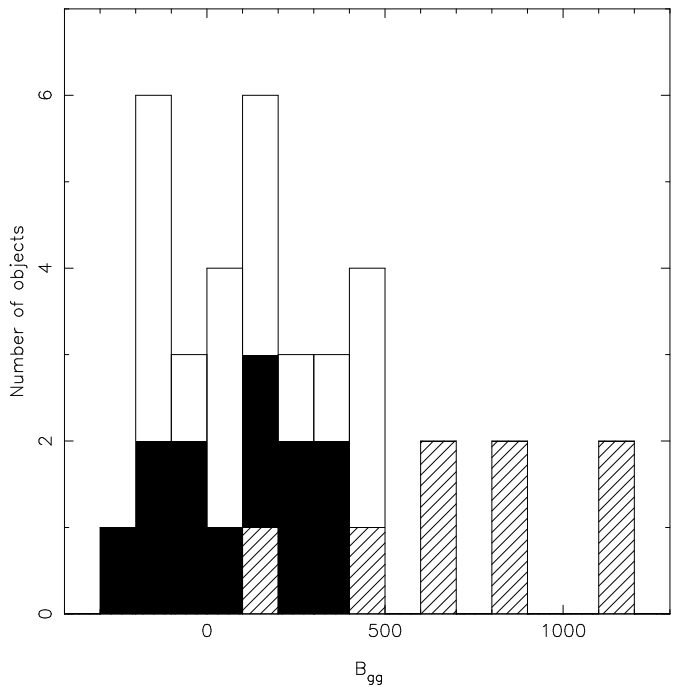}
\caption{Histogram of $B_{\rm gg}$ values for the $0.15 < z < 0.4$
  objects from the HESR 3CR sample. Empty boxes indicate narrow-line
  objects, filled boxes broad-line objects (including quasars) and
  hatched boxes low-excitation objects.}
\label{bgghisto}
\end{figure}

\section{Data and analysis}

The data for 3CR objects, including the emission-line classifications,
are taken from HESR, with emission-line classifications supplemented
by data from the literature (Table \ref{emlin}). HESR took most of
their emission-line classifications from Jackson \& Rawlings (1997).
Lacking new spectroscopic observations, I have classified objects
\new{that were not classified in Jackson \& Rawlings} mostly on the
basis of qualitative statements in the literature, such as whether
they show `strong' or `weak' emission lines and what species are
present in the spectrum. These classifications must be regarded as
best guesses only, but, as discussed by Hardcastle \etal\ (1998), the
approach tends to give the same results as a more quantitative
determination. \new{To check this quantitatively, I examined all
galaxies that are classified both in Jackson \& Rawlings and in
Spinrad \etal\ (1985), comparing Spinrad \etal 's qualitative spectral
classification with the Jackson \& Rawlings class. The two papers were
taken to agree if an object classed as Spinrad \etal\ as `WE' or `ABS'
was classed by Jackson \& Rawlings as a low-excitation object, or if
an object classed by Spinrad \etal\ as `SE' was classed by Jackson \&
Rawlings as a high-excitation object or `weak quasar' (effectively a
broad-line radio galaxy). Of the 97 overlapping objects, the two
papers disagreed on only 8. This gives us some confidence that a
qualitative approach will produce acceptable results.} One object,
3C\,142.1, remains unclassified, but lies outside the redshift range
of particular interest in any case. Following HESR, 3C\,275 and
3C\,435A were omitted from the analysis. The data for HESR's
comparison sample of radio-loud quasars were taken directly from YE93;
HESR's $B_{\rm gg}$ values were taken in preference to YE93's where
the sample had objects in common.

Fig. \ref{bgghisto} shows the distribution of $B_{\rm gg}$ values for
the objects in the 3CR sample with $0.15 < z < 0.4$, classified by
emission-line properties. This may be compared with the left-hand
panel of HESR's figure 6. It is clear that the low-excitation objects
have a different distribution of $B_{\rm gg}$ from the high-excitation
radio galaxies and quasars; in fact, all the 3CR objects in the
richest environments found by HESR are LERG.

HESR used a Kolmogorov-Smirnov test to investigate the differences in
$B_{\rm gg}$ distribution between the radio galaxies and quasars in
their sample with $0.15 < z < 0.4$. To compare with HESR's approach, I
use a quasar sample consisting of 21 YE93 quasars (the 24 objects in
the redshift range, omitting the 3 quasars in the HESR 3CR sample)
plus the 5 quasars from the 3CR sample, giving 26 quasars in all. In
the HESR redshift range there are 31 radio galaxies (including BLRG
and N-galaxies; here I class these with the NLRG for consistency with
HESR). The K-S test finds only a 5\% probability that these two
samples are drawn from the same parent distribution, consistent with
HESR's value of 3\%. But if the 8 LERG are excluded from the radio
galaxy sample, the K-S probability that the two are from the same
parent population rises to 50\%; there is no significant difference
between the distributions of radio galaxies and quasars in these
samples in this redshift range.

We can restrict ourselves to the 3CR-derived sample with $0.15 < z <
0.4$ in order to test unified models without including the
lower-radio-power YE93 quasars. \new{Here, in order to obtain a
  sufficiently large sample of sources that make a small angle to the
  line of sight to apply statistical tests, I include the BLRG with
  the quasars; this is a valid procedure whether or not some of the
  BLRG are intermediate-angle objects.} In this sample, the $B_{\rm
  gg}$ distribution of the 12 broad-line objects (quasars and BLRG)
has only a 5\% probability of being drawn from the same population as
the 24 LERG and NLRG combined (cf.\ Fig.\ \ref{bgghisto}), but a 72\%
probability of being drawn from the same population as the 16 NLRG;
thus, if we neglect the difference between LERG and NLRG, there is a
problem for low-redshift unification as described by Hardcastle \etal\
(1998), while if we remove the LERG there is no problem. On the other
hand, the $B_{\rm gg}$ distribution of the LERG with $0.15 < z < 0.4$
has a 0.04\% probability of being drawn from the same population as
the other 3CR objects, a highly significant result. To summarize, all
the differences between the distributions of radio galaxies and
quasars/broad-line objects in the redshift range $0.15 < z < 0.4$
become statistically insignificant if the LERG are excluded from the
radio galaxy sample; the LERG have significantly different properties
from the other objects.

\section{Discussion and conclusions}
\label{discussion}

The key result of this work is that 3CR LERG lie in significantly
richer environments than the HERG/quasar population at comparable
redshifts. This strong difference between the environmental properties
of LERG and HERG supports suggestions by Hardcastle \etal\ (1998) and
Hardcastle \& Worrall (1999) that the 3CR LERG population is
physically different from the HERG. If we believe that the
emission-line properties of the radio source reflect the power of the
jet (e.g. Rawlings \& Saunders 1991) then a plausible explanation is
that at least some LERG are intrinsically low-jet-power radio galaxies
which owe their comparatively high overall radio luminosity and their
other properties (small size, dissipative jets, relaxed appearance --
cf. Hardcastle \etal\ 1998, Harvanek \& Stocke 2002) to
a rich environment. This recalls, and may be related to, the arguments
of Barthel \& Arnaud (1996).

The fact that the narrow-line objects and quasars have similar $B_{\rm
gg}$ distributions, when the LERG are excluded, supports the version
of the orientation-based unified model for radio galaxies and quasars
outlined in section \ref{unif}.

\begin{acknowledgements}
I thank the Royal Society for a Research Fellowship, and Michael
Harvanek for discussions of $B_{\rm gg}$ values for radio galaxies in
the mid-1990s which helped to alert me to the interesting properties
of LERGs. I am grateful to Michael Harvanek, John Stocke and Judith
Croston for their helpful comments on an earlier draft of this paper,
and to Philip Best for a rapid and constructive referee's report.
\end{acknowledgements}

\end{document}